\documentclass[pra,aps,twocolumn,amsfonts,showpacs,superscriptaddress,floatfix]{revtex4-1}

\usepackage{epsfig} 
\usepackage{psfrag} 
\usepackage{amsmath}
\usepackage{amssymb} 
\usepackage{color} 
\usepackage{graphicx}
\usepackage{subfigure}

\newcommand{\bra}[1]{\ensuremath{\left\langle #1 \right|}}
\newcommand{\ket}[1]{\ensuremath{\left| #1 \right\rangle}}

\newcommand{\expec}[1]{\ensuremath{\langle #1\rangle}}

\begin{document}

\title{Magnetic Properties of the Second Mott Lobe in Pairing
  Hamiltonians}

\author{M. J. Bhaseen} \affiliation{University of Cambridge, 
Cavendish Laboratory, Cambridge, CB3 0HE, UK.}
\author{S. Ejima} 
\affiliation{Institut f{\"u}r Physik, 
Ernst-Moritz-Arndt-Universit{\"a}t Greifswald, 17489 Greifswald, Germany.}  
\author{M. Hohenadler} 
\affiliation{Institut f\"ur Theoretische Physik und Astrophysik, 
Universit\"at W\"urzburg, Germany.}
\author{A. O. Silver} \affiliation{University of Cambridge, 
Cavendish Laboratory, Cambridge, CB3 0HE, UK.}
\author{F. H. L. Essler} 
\affiliation{The Rudolf Peierls Centre for Theoretical Physics, 
University of Oxford, Oxford, OX1 3NP, UK.}
\author{H. Fehske} \affiliation{Institut f{\"u}r Physik,
Ernst-Moritz-Arndt-Universit{\"a}t Greifswald, 17489 Greifswald, Germany.}  
\author{B. D. Simons} \affiliation{University of
Cambridge, Cavendish Laboratory, Cambridge, CB3 0HE, UK.}

\begin{abstract}
  We explore the Mott insulating state of single-band bosonic pairing
  Hamiltonians using analytical approaches and large scale density
  matrix renormalization group calculations. We focus on the second
  Mott lobe which exhibits a magnetic quantum phase transition in the
  Ising universality class. We use this feature to discuss the
  behavior of a range of physical observables within the framework of
  the 1D quantum Ising model and the strongly anisotropic Heisenberg
  model. This includes the properties of local expectation values and
  correlation functions both at and away from criticality. Depending
  on the microscopic interactions it is possible to achieve either
  antiferromagnetic or ferromagnetic exchange interactions and we
  highlight the possibility of observing the ${\rm E}_8$ mass spectrum
  for the critical Ising model in a longitudinal magnetic field.
\end{abstract}

\date{\today}


\maketitle

\section{Introduction}

The observation of Bose--Einstein condensation (BEC) in dilute alkali
gases \cite{Anderson:BEC,Ketterle:BEC} has led to a wealth of activity
linking ultra cold atoms research and condensed matter physics. The
precise control over atomic interactions, and the use of optical
lattices, offers valuable insights into the effects of strong
correlations \cite{Jaksch:CBAOL}.  This is exemplified by experiments
on the Bose--Hubbard model which reveal the quantum phase transition
from a superfluid (SF) to a Mott insulator (MI) as the depth of the
optical lattice is increased \cite{Greiner:SI,Fisher:Bosonloc}.
Motivated by these advances, recent attention has been directed
towards a variety of multicomponent systems, including spinor
condensates
\cite{Stenger:Spindomains,Misener:Obs,Stamper:Tunneling,Leanhardt:Coreless,Ohmi:Internal,Ho:Spinor,Demler:Spinor,Imambekov:Spinone,Lamacraft:QQSC,Mukerjee:Dynamical,Essler:Spinor,Ueda:Spinor}
and atomic mixtures
\cite{Thalhammer:Double,Cazalilla:Instab,Kuklov:Commens,Kleine:Spincharge,*Kleine:Excitations,Soyler:Twocomp,Guglielmino:BB}. The
possibility of novel behavior is greatly enhanced in the presence of
these additional ``spin'' degrees of freedom, and routes to quantum
magnetism have been proposed by exploiting internal hyperfine states
\cite{Duan:Control} and using different atomic species
\cite{Altman:Twocomp,Hubener:Magnetic,Powell:2BH,Capo:Critical}. The
decoupling of the electronic and nuclear spins in alkaline earth atoms
has also been suggested as a way to realize quantum spin liquids and
exotic magnetism based on the ${\rm SU}({\rm N})$ groups
\cite{Gorshkov:Two,Hermele:CSL,Cazalilla:SU}.

In recent work some of the present authors suggested using
atom--molecule mixtures as a route to the paradigmatic quantum Ising
model \cite{Bhaseen:Feshising,Hohenadler:QPT}. Motivated by studies of
the BEC-BCS transition for bosons, both in the continuum limit
\cite{Rad:Atmol,Romans:QPT,Radzi:Resonant} and on the lattice
\cite{Dickerscheid:Feshbach,Sengupta:Feshbach,Rousseau:Fesh,Rousseau:Mixtures},
we investigated the rich phase diagram of bosons interacting via
Feshbach resonant pairing interactions in an optical
lattice. Combining exact diagonalization (ED) on small systems with
analytical predictions based on the strong coupling expansion, we
provided evidence for an Ising quantum phase transition in the second
Mott lobe. In contrast to previous numerical studies which advocated
the presence of super-Mott behavior
\cite{Rousseau:Fesh,Rousseau:Mixtures}, the Ising spectral gap
indicates the absence of low lying superfluid excitations deep within
the Mott phase \cite{Bhaseen:Feshising,Hohenadler:QPT}.  This
conclusion gained further support in a recent comment
\cite{Eckholt:Comment} which shows the presence of exponential decay
in the connected correlation functions. A detailed discussion of the
superfluid properties \cite{Rad:Atmol,Romans:QPT,Radzi:Resonant} in
one dimension (1D) was also provided in Ref.~\cite{Ejima:ID} using
large scale density matrix renomalization group (DMRG)
\cite{White:DMRG,*White:Algorithms} and field theory techniques.

In view of the broad interest in these systems, and the technical
issues surrounding Refs.~\cite{Rousseau:Fesh,Rousseau:Mixtures}, we
discuss the properties of the second Mott lobe in detail using
analytical arguments and DMRG. To aid the comparison with these
previous works we focus primarily on the homonuclear case with a
single species of bosonic atom. We also begin by restricting the local
Hilbert space for the atomic and molecular occupations
\cite{Rousseau:Mixtures}.  In this reduced setting we determine both
the locus of the antiferromagnetic Ising transition and the onset of
superfluidity.  We also investigate the atomic and molecular
correlation functions within the MI and compare directly to the
predictions of the quantum Ising model. As advocated in
Refs.~\cite{Bhaseen:Feshising,Hohenadler:QPT} this provides a simple
and intuitive framework in which to discuss the absence of super-Mott
behavior \cite{Rousseau:Fesh,Rousseau:Mixtures,Eckholt:Comment}.

Going beyond the restricted Hilbert space description we show that the
same ideas apply. Interestingly, by tuning the microscopic
interactions it is also possible to change the sign of the Ising
exchange interaction from antiferromagnetic (AFM) to ferromagnetic
(FM).  In general these Ising Hamiltonians also contain an effective
magnetic field \cite{Bhaseen:Feshising,Hohenadler:QPT} and in 1D this
will act as a confinement potential for the zero field excitations of
the FM chain \cite{Mccoy:Breakup,Bhaseen:Aspects}.  This suggests the
possibility of observing the non-trivial ${\rm E}_8$ mass spectrum of
``meson" bound states for the critical FM Ising model in a
longitudinal field
\cite{Zamolodchikov:Integrals,Coldea:E8,Moore:E8}. As an extension of
these results we also provide the magnetic Hamiltonian for Bose--Fermi
mixtures.

The layout of this paper is as follows. In Secs.~\ref{Sect:Model} and
\ref{Sect:MI2} we provide an introduction to the bosonic Feshbach
Hamiltonian and discuss the mapping to the quantum Ising model.  In
Sec.~\ref{Sect:PD} we present a cross section of the phase diagram
obtained by DMRG which displays both the Ising quantum phase
transition and the onset of superfluidity. We confirm the Ising
behavior within the Mott phase using results for the excitation gap
and the entanglement entropy. In Sec.~\ref{Sect:XY} we investigate the
role of higher order terms in the strong coupling expansion. We
discuss their impact on the non-universal properties of the phase
diagram such as the curvature of the Ising phase boundary. In
Sec.~\ref{Sect:Local} we examine the local expectation values within
the Mott phase at and away from criticality.  We move on to
correlation functions in Secs.~\ref{Sect:Corr} and \ref{Sect:Atmol}
and discuss their relation to the Ising model.  In
Secs.~\ref{Sect:Soft} and \ref{Sect:FM} we turn our attention to
softcore bosons and demonstrate the existence of both
  antiferromagnetic and ferromagnetic Ising transitions.  We comment
on the closely related fermionic problem in Appendix \ref{Sect:Fermi}.
We conclude in Sec.~\ref{Sect:Conc} and provide perspectives for
further research.

\section{Model}
\label{Sect:Model}

We consider the Hamiltonian
\cite{Dickerscheid:Feshbach,Sengupta:Feshbach,Rousseau:Fesh,Rousseau:Mixtures}
\begin{eqnarray}
H & = & \sum_{i\alpha} \epsilon_\alpha n_{i\alpha}-\sum_{\langle ij\rangle}\sum_\alpha 
t_\alpha(b_{i\alpha}^\dagger b_{j\alpha}+{\rm H.c.}) \nonumber \\
& & \hspace{1cm} +\sum_{i\alpha\alpha^\prime}
\frac{U_{\alpha\alpha^\prime}}{2}:n_{i\alpha}n_{i\alpha^\prime}:+H_{\rm F},
\label{atmolham}
\end{eqnarray}
describing bosons, $b_{i\alpha}$ hopping on a lattice with sites $i$
where $\alpha=a,m$ labels atoms and molecules and
$n_{i\alpha}=b_{i\alpha}^\dagger b_{i\alpha}$. Here, $\epsilon_\alpha$
are on-site potentials, $t_\alpha$ are hopping parameters, $\langle
ij\rangle$ denotes summation over nearest neighbor bonds and
$U_{\alpha\alpha^\prime}$ are interactions. We use normal ordering
symbols to indicate ${:n_{i\alpha}n_{i\alpha^\prime}:} =
n_{i\alpha}(n_{i\alpha}-1)$ for like species and
${:n_{i\alpha}n_{i\alpha^\prime}:} = n_{i\alpha}n_{i\alpha^\prime}$
for distinct species.  For simplicity we assume that molecules are
formed by $s$-wave Feshbach resonant interactions
\begin{equation}
H_{\rm F}=g\sum_i(m_i^\dagger a_i a_i+{\rm H.c.}),
\label{HF}
\end{equation}
where $m_i\equiv b_{im}$ and $a_i\equiv b_{ia}$; for recent work on
the $p$-wave problem see Ref.~\cite{Radzi:Spinor}. An important
feature of the Feshbach interaction (\ref{HF}) is that atoms and
molecules are not separately conserved. However, the total, $N_{\rm
  T}\equiv \sum_i(n_{ia}+2n_{im})$ is preserved. One may therefore
work in the canonical ensemble with $\rho_{\rm T}\equiv N_{\rm T}/L$
held fixed, where $L$ is the number of lattice sites. In order to make
contact with the previous literature
\cite{Rousseau:Fesh,Rousseau:Mixtures,Eckholt:Comment} we consider
only a single-band description in Eq.~(\ref{atmolham}); for a
discussion of higher band effects see
Refs.~\cite{Diener:comment,Dickerscheid:Reply,Buchler:Micro}. As
reported elsewhere the pairing Hamiltonian (\ref{atmolham}) has a rich
phase diagram exhibiting both MI and SF phases
\cite{Rad:Atmol,Romans:QPT,Radzi:Resonant,Dickerscheid:Feshbach,Sengupta:Feshbach,Rousseau:Fesh,Rousseau:Mixtures,Bhaseen:Feshising,Hohenadler:QPT,Ejima:ID}. 
Most notably, the system displays a discrete ${\mathbb Z}_2$ symmetry
breaking transition \cite{Rad:Atmol,Romans:QPT,Radzi:Resonant} between
a paired molecular condensate (MC) and an atomic plus molecular
condensate (AC+MC) phase \cite{Ejima:ID}; for closely related
transitions in other models see also
Refs. \cite{Daley:Three,*Daley:Threeerratum,*Diehl:Observability,Diehl:QFTI,*Diehl:QFTII,Wu:Competing,*Lecheminant:Confinement,Capponi:Confinement,*Roux:Spin,Bonnes:Pair}.
Here, our main focus will be on the MI phase. In particular, we shed
further light on the magnetic characteristics of the second Mott lobe
\cite{Bhaseen:Feshising,Hohenadler:QPT}.  We also discuss the
connection to Ref.~\cite{Eckholt:Comment} and earlier numerical
studies \cite{Rousseau:Fesh,Rousseau:Mixtures}. In order to facilitate
our numerical simulations we consider the restricted Hamiltonian
\begin{equation}
H_{\mathcal P}={\mathcal P}H{\mathcal P},
\label{HR}
\end{equation}
where the projection operator ${\mathcal P}$ projects on to the
restricted local Hilbert space with a maximum of $r_a$ atoms and $r_m$
molecules per site.  We begin our discussion in Sec.~\ref{Sect:MI2} by
setting $r_a=2$ and $r_m=1$ as used in Ref.~\cite{Rousseau:Mixtures}.
As we will see, the essential characteristic features of the phase
diagram are readily gleaned from this limiting case. We move on to the
more general problem with canonical softcore bosons in
Secs.~\ref{Sect:Soft} and \ref{Sect:FM}.  Throughout this manuscript
we use the value of $U_{am}$ to set the overall energy scale; in
Figs.~\ref{Fig:PD}, \ref{Fig:AFMgap}, \ref{Fig:Entropy} 
and Figs.~\ref{Fig:Localmag}, \ref{Fig:Mottcorr}, \ref{Fig:DMRGcollapse1} 
we set $U_{am}=1$, and in the remaining figures we
set $U_{am}=4$ in order to descend deeper in to the Mott phase.

\section{Second Mott Lobe} 
\label{Sect:MI2}

\begin{figure}
  \includegraphics[clip=true,width=\columnwidth]{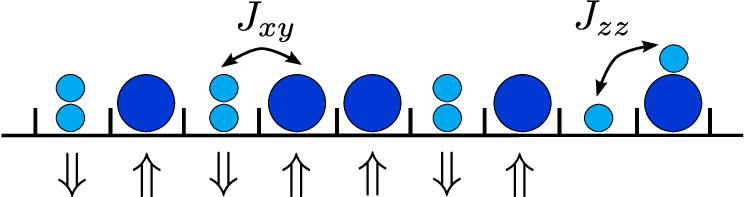}
  \caption{(color online). Depiction of the second Mott lobe of the
    Hamiltonian (\ref{atmolham}) with $\rho_{\rm T}=2$ showing either
    a pair of atoms or a molecule on each site. These may be regarded
    as an effective spin-down ($\Downarrow$) and spin-up ($\Uparrow$)
    respectively. Second order virtual hopping processes lead to
    $J_{zz}$ interactions and an effective Ising model. XY exchange,
    $J_{xy}$, occurs at third order and involves interchanging a pair
    of atoms and a molecule. The Feshbach term (\ref{HF}) converts a
    pair of atoms ($\Downarrow$) into a molecule ($\Uparrow$) and
    therefore acts like a transverse field.}
\label{Fig:Secondlobe}
\end{figure}

A convenient way to describe the second Mott lobe with $\rho_{\rm
  T}=2$ is via an effective spin model derived within the strong
coupling expansion; see Fig.~\ref{Fig:Secondlobe}.
Introducing effective spins 
$|\!\Downarrow\rangle\equiv |2;0\rangle/\sqrt{2}$ and
$|\!\Uparrow\rangle\equiv |0;1\rangle$ in the occupation basis
$|n_a;n_m\rangle$
one obtains the effective spin-$1/2$ quantum Ising model
\cite{Bhaseen:Feshising,Hohenadler:QPT}
\begin{equation}
H\simeq J_{zz}\sum_{i}S^z_i S^z_{i+1}+h\sum_iS^z_i+\Gamma\sum_iS^x_i+C
+{\mathcal O}(t^3),
\label{Ising}
\end{equation}
where we work to second order in the hopping parameters.  The spin
operators are given by
\begin{equation}
S^+=\frac{m^\dagger a a}{\sqrt{2}}, \quad 
S^-=\frac{a^\dagger a^\dagger m}{\sqrt{2}}, \quad 
S^z=\frac{(n_m-n_a/2)}{2},
\label{spindef}
\end{equation}
and $S^\pm \equiv S^x\pm i S^y$. The factors of $\sqrt{2}$ arise from
the action of the Bose operators on the basis states.  Equivalently,
since $n_a+2n_m=2$, one may also write $S^z=(1-n_a)/2$ or
$S^z=(2n_m-1)/2$. As indicated in Fig.~\ref{Fig:Secondlobe}, the first
term in Eq.~(\ref{Ising}) arises from the virtual hopping processes of
atoms and molecules on to neighboring sites and corresponds to an
effective magnetic exchange interaction, $J_{zz}$. The second term in
Eq.~(\ref{Ising}) reflects the energetic detuning between atoms and
molecules and corresponds to an effective longitudinal magnetic field,
$h$.  The third term in Eq.~(\ref{Ising}) corresponds directly to the
Feshbach term in Eq.~(\ref{HF}). Indeed, it is readily seen from the
definition (\ref{spindef}) that $S^+$ converts two atoms
($\Downarrow$) into a molecule ($\Uparrow$) and therefore acts like a
spin raising operator. It follows that the Feshbach term (\ref{HF})
acts like a transverse field in the $x$-direction with
\begin{equation}
\Gamma\equiv 2g\sqrt{2}. 
\end{equation}
The overall structure of the magnetic Hamiltonian (\ref{Ising}) is
generic to the second Mott lobe with $\rho_{\rm T}=2$. However, the
coefficients depend on the specific Hilbert space restriction. For the
Hamiltonian (\ref{HR}) with restriction parameters $r_a=2$ and $r_m=1$
one obtains \cite{Bhaseen:Feshising,Hohenadler:QPT}
\begin{equation}
J_{zz}=\frac{4t_a^2}{U_{am}-U_{aa}}+\frac{t_m^2}{U_{am}},
\label{Isingj}
\end{equation} 
and
\begin{equation}
h=\epsilon_m-2\epsilon_a-U_{aa}.
\label{Isingh}
\end{equation}
The constant offset is given by
\begin{equation}
C=L\left(\frac{\epsilon_m}{2}+\epsilon_a+\frac{U_{aa}}{2}
-\frac{zJ_{zz}}{8}\right),
\label{Isingc}
\end{equation}
where $z$ is the cubic lattice coordination and $z=2$ in 1D. In
general there will also be additional contributions to these
coefficients arising from higher order terms in the strong coupling
expansion. As we will discuss in more detail below similar results
also hold for canonical softcore atoms and molecules but with modified
coefficients. This is due to the presence of additional intermediate
states that are explored in the virtual hopping
processes. Nonetheless, this truncation of the Hilbert space is useful
for initial numerical simulations, and is also used in
Ref. \cite{Rousseau:Mixtures}. In addition the principal features of
the magnetic description are more readily exposed. We will return to
canonical softcore bosons in Secs.~\ref{Sect:Soft} and \ref{Sect:FM}.

\section{Phase Diagram and the Antiferromagnetic 
Ising Transition}
\label{Sect:PD}

In Refs.~\cite{Bhaseen:Feshising,Hohenadler:QPT} we provided a variety
of evidence for an Ising quantum phase transition in the closely
related heteronuclear generalization of Hamiltonian
(\ref{atmolham}). However, due to the small system sizes accessible by
exact diagonalization, a complete elucidation of the phase diagram was
not possible. In addition correlation functions were out of reach.  In
particular, it was not possible to track the Ising quantum phase
transition throughout the second Mott lobe, or to accurately delimit
the onset of superfluidity. We address this situation for the
homonuclear case by using DMRG on large systems.  We keep up to 3000
density matrix states so that the discarded weight is less than
$10^{-10}$. Following Ref.~\cite{Rousseau:Mixtures} we implement the
model (\ref{HR}) as a two-leg ladder system, where the atoms and
molecules reside on opposite legs of the ladder. In this
representation the Feshbach term (\ref{HF}) corresponds to hopping
along the rungs, and extreme care must be taken for small values of
$g$.
\begin{figure}[!h]
 \includegraphics[clip=true,width=\columnwidth]{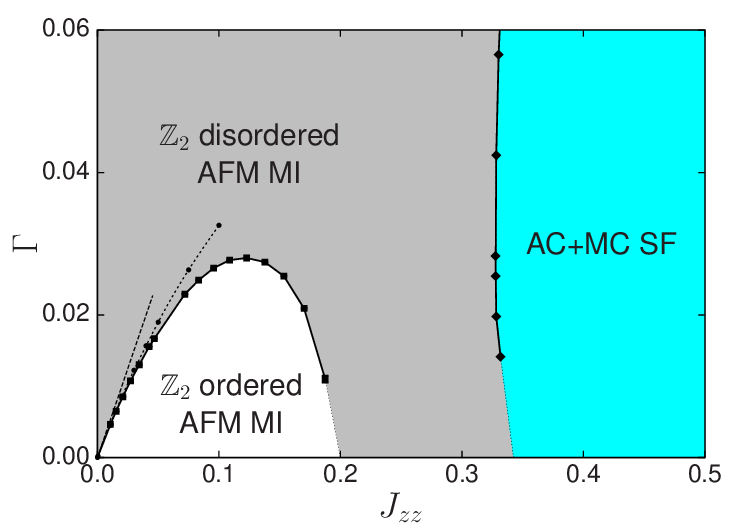}
 \caption{(color online). Phase diagram of the 1D Hamiltonian (\ref{HR}) with
   $\rho_{\rm T}=2$ obtained by DMRG with up to $L=256$ sites.  We
   restrict the local Hilbert space to a maximum of one molecule and
   two atoms per site with $r_m=1$ and $r_a=2$, and set
   $\epsilon_a=\epsilon_m=0$, $U_{aa}=0$, $U_{am}=1$, corresponding to
   $h=0$. We set $t_a=2t_m=t$ and vary $t$ to obtain different values
   of $J_{zz}=17t^2/4$ corresponding to antiferromagnetic (AFM)
   exchange.  This cross section shows a ${\mathbb Z}_2$ ordered AFM
   MI, a ${\mathbb Z}_2$ disordered AFM MI, and an AC+MC SF. The
   filled squares are obtained from the vanishing of the first
   excitation gap, $\Delta_1\equiv E_1-E_0$ corresponding to an Ising
   quantum phase transition within the MI; the solid line is a spline
   fit which is extrapolated down to $\Gamma=0$. 
The dashed line corresponds to the strong coupling result,
   $\Gamma_{\rm c}=J_{zz}/2$ which follows from the Hamiltonian
   (\ref{Ising}) with $h=0$. The dotted line is obtained by ED of
   Hamiltonian (\ref{Ising}) including the subleading XY exchange
   terms, $J_{xy}=-0.46\,J_{zz}^{3/2}$ which arise at higher order in
   the strong coupling expansion. The diamonds are obtained from the
   vanishing of the two-particle gap, $E_{2g}\equiv \mu_{+}-\mu_{-}$
   where $\mu_{\pm}=\pm[E_0(L,N_{\rm T}\pm 2)-E_0(L,N_{\rm T})]$ and
   up to $L=128$, indicating the onset of an AC+MC superfluid.}
\label{Fig:PD}
\end{figure}
In Fig.~\ref{Fig:PD} we present a cross section of the phase diagram
for the Hamiltonian (\ref{HR}) with $r_a=2$ and $r_m=1$. In this
  approach the Hilbert space restriction parameters $r_\alpha$ and the
  interactions $U_{\alpha\alpha^\prime}$ are treated
  independently. For ease of exposition we begin by setting
  $\epsilon_a=\epsilon_m=U_{aa}=0$ and $U_{am}=1$, where our choice of
  parameters is motivated by the simplest case with zero magnetic
  field as given by Eq.~(\ref{Isingh}).  In addition, for small
  hopping parameters $t_\alpha$, and large $U_{am}$, one may set
  $U_{aa}=0$ in Eq. (\ref{Isingj}) without loss of generality or
  conflict with the strong coupling expansion. We will incorporate the
  effects of finite $U_{aa}$ in our subsequent discussion.

\begin{figure}[!h]
  \includegraphics[clip=true,width=3.2in]{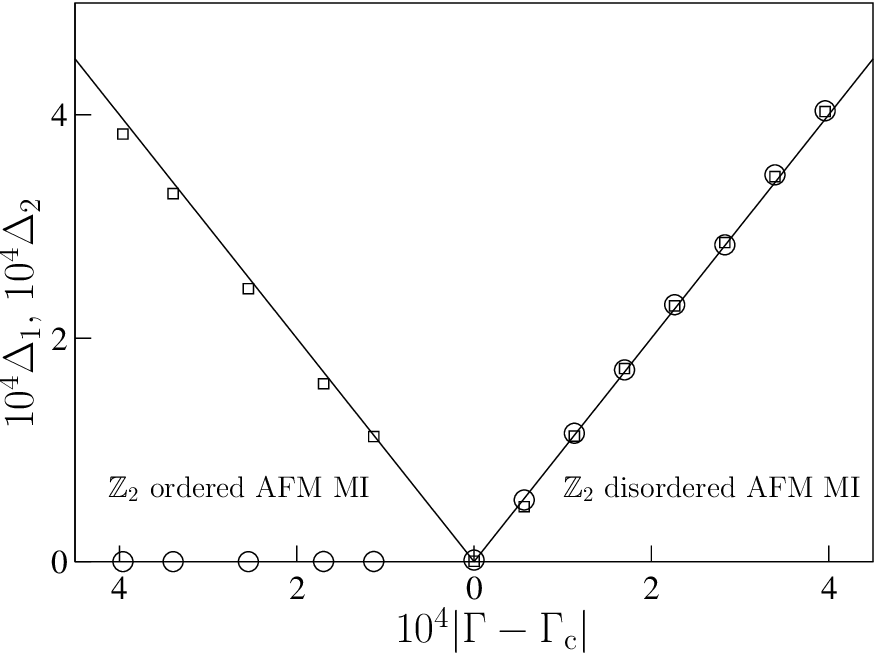}
  \caption{DMRG results for the excitation gaps, $\Delta_1\equiv
    E_1-E_0$ (circles) and $\Delta_2\equiv E_2-E_0$ (squares), for
    the bosonic Hamiltonian (\ref{HR}) with open boundaries and
    up to $L=256$ sites. We pass
    through the Ising quantum phase transition shown in
    Fig.~\ref{Fig:PD} with $t=0.05$ and $J_{zz}\approx 0.01$. The
    transition occurs at $\Gamma_{\rm c}\approx 4.6\times 10^{-3}$ which is
    slightly below the strong coupling result,
    $\Gamma_{\rm c}=J_{zz}/2\approx 5.3\times 10^{-3}$. This reflects the
    curvature of the Ising transition shown in Fig.~\ref{Fig:PD}. The
    solid lines indicate the linear gap,
    $\Delta=|\Gamma-\Gamma_{\rm c}|$, corresponding to the Ising critical
    exponent, $\nu=1$.}
\label{Fig:AFMgap}
\end{figure}

\begin{figure}[!h]
\includegraphics[clip=true,width=3.2in]{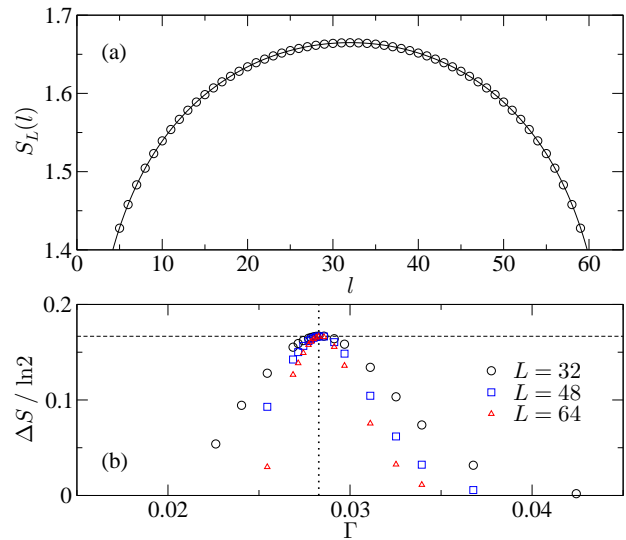}
\caption{(a) DMRG results for the entanglement entropy, $S_L(l)$, at
  the Ising transition within the MI for a subsystem of length $l$ in
  an periodic chain with $L=64$.  We use the same parameters as in
  Fig.~\ref{Fig:PD} and set $t=0.17$ corresponding to
  $J_{zz}\approx 0.12$ and $\Gamma_c\approx 0.028$.  The solid line
  is a fit to Eq.~(\ref{SLlpbc}). The extracted value of the central
  charge $c=0.50(2)$ is consistent with a magnetic transition in the
  Ising universality class. (b) The entanglement entropy difference
  $\Delta S(L)$ for $t=0.17$ shows a peak at the Ising transition. The
  peak height corresponds to $c=1/2$ as indicated by the dashed
  lines.}
\label{Fig:Entropy}
\end{figure}

  The phase diagram in Fig.~\ref{Fig:PD} contains three distinct
  phases.  A ${\mathbb Z}_2$ disordered MI with vanishing staggered
  magnetization $\sum_i(-1)^i\langle S_i^z\rangle/L=0$, a ${\mathbb
    Z}_2$ ordered MI with a finite staggered magnetization and long
  range antiferromagnetic correlations, and an AC+MC superfluid with
  both atomic and molecular power law superfluidity. The additional MC
  phase with only molecular superfluidity is absent in this cross
  section of the phase diagram due to our choice of parameters; for
  more details of the superfluid phases see Ref.~\cite{Ejima:ID}.  In
  this manuscript our main focus is on the MI phase. In particular, we
  see that the magnetic phase boundary bends over quite considerably
  due to higher order terms in the strong coupling expansion.
  Nonetheless, the quantum phase transition remains in the Ising
  universality class.  For example, the Ising character of this
  transition is supported by Fig.~\ref{Fig:AFMgap} which shows the
  characteristic linear variation of the excitation gap
  \cite{Sachdev:QPT}
\begin{equation}
\Delta=|\Gamma-\Gamma_{\rm c}|,
\end{equation}
on passing through the transition in the scaling regime.
The Ising character is also confirmed by DMRG results for the entanglement
entropy as shown in Fig.~\ref{Fig:Entropy}.
For a block of length $l$ in a system of length $L$, the von Neumann
entropy is given by $S_L(l)=-{\rm Tr}_l\left(\rho_l\ln\rho_l\right)$,
where $\rho_l={\rm Tr}_{L-l}(\rho)$ is the reduced density matrix. In
a critical system with periodic boundaries one obtains
\cite{Holzey:Entropy,Korepin:Entropy,Calabrese:Entanglement,Legeza:Entropic}
\begin{equation}
  S_L(l)=\frac{c}{3}\ln \left[\frac{2L}{\pi}\sin\left(\frac{\pi l}{L}\right)\right]+s_1,
\label{SLlpbc}
\end{equation}
where $s_1$ is a non-universal constant and $c$ is the central charge.
As shown in Fig.~\ref{Fig:Entropy}(a) the results are in
excellent agreement with an Ising quantum phase transition with
$c=1/2$. In particular, this continues to hold further out in the
  Mott phase where the strong coupling analysis no longer strictly
  applies.  Further evidence for this may be seen from the
entanglement entropy difference \cite{LK:Spreading,Ejima:ID}
\begin{equation}
\Delta S(L)\equiv S_L(L/2)-S_{L/2}(L/4)
\label{deltas}
\end{equation}
which exhibits a peak on transiting through the magnetic transition as
shown in Fig.~\ref{Fig:Entropy}(b). At criticality $\Delta
S=\frac{c}{3}\ln 2+\dots$
\cite{Holzey:Entropy,Korepin:Entropy,Calabrese:Entanglement,Legeza:Entropic}
and the peak height is in good agreement with $c=1/2$.  
It is interesting to note
that a naive spline fit to the currently available DMRG data for the
Ising phase boundary shown in Fig.~\ref{Fig:PD} terminates within the
MI. This tentatively suggests that a magnetic transition may exist
even in the absence of the Feshbach term (\ref{HF}).  Unfortunately,
it is difficult to gain a quantitative handle on the character of this
feature, which occurs for intermediate hopping strengths and vanishing
Feshbach coupling, using our current implementation of DMRG. Moreover,
this $\Gamma=0$ limit differs somewhat from other investigations
of the two-component Bose--Hubbard model \cite{Kuklov:Counter,Hu:Counter}, 
since we keep the total
density, $\rho_{\rm T}=\sum_i(n_{ia}+2n_{im})/L=2$ held fixed, with
two atoms and one molecule per site. It would be instructive to
explore this limit in future work.

\section{Higher Order Contributions in the Strong Coupling Expansion}
\label{Sect:XY}

In the discussion above we have presented a variety of large scale
DMRG results in favor of an Ising quantum phase transition taking
place within the second Mott lobe
\cite{Bhaseen:Feshising,Hohenadler:QPT}. In particular we have argued
that the Ising character of the quantum phase transition persists
throughout an extended region of the phase diagram as shown in
Fig.~\ref{Fig:PD}. Nonetheless, it is important to bear in mind that
the explicit Ising Hamiltonian given in Eq.~(\ref{Ising}) is derived
by means of the strong coupling, $t/U$ expansion. It is therefore
tailored towards a quantitative description of the bosonic Hamiltonian
(\ref{atmolham}) deep within the Mott lobe, as supported by our DMRG
results.  At larger values of the hopping parameters departures from
the strong coupling result (\ref{Ising}) are to be expected. This is
evident from the deviation of the phase boundary from the asymptotic
result, $\Gamma_{\rm c}=J_{zz}/2$ as shown in
Fig.~\ref{Fig:PD}. Carrying out the strong coupling expansion to third
order in the hopping parameters one must supplement the Hamiltonian
(\ref{Ising}) with the additional XY exchange terms:
\begin{equation}
\Delta H= \frac{J_{xy}}{2}\sum_{i}\left(S_i^+S_{i+1}^-+S_i^-S_{i+1}^+\right),
\label{hxy}
\end{equation}
where $S^\pm \equiv S^x\pm iS^y$, and 
\begin{equation}
  J_{xy}=-\frac{4t_a^2t_m}{U_{am}-U_{aa}}\left(\frac{1}{U_{am}}
    +\frac{1}{U_{am}-U_{aa}}\right).
\label{XY}
\end{equation}
One thus obtains a strongly anisotropic XXZ Heisenberg Hamiltonian in
both a longitudinal and transverse field. For the parameters chosen in
Fig.~\ref{Fig:PD}, $J_{zz}=17t^2/4$ and $J_{xy}=-4t^3$. This yields
the XXZ anisotropy parameter, $\delta\equiv J_{xy}/J_{zz}=-16t/17$
corresponding to $\delta\approx -0.46 \sqrt{J_{zz}}$. Within the MI
shown in Fig.~\ref{Fig:PD} this gives $-0.27 \lesssim \delta <0$. The
system remains in the strongly anisotropic regime and Ising
criticality is expected to persist throughout this cross section of
the MI.  Nonetheless, the XY contributions (\ref{hxy}) modify the
location of the Ising quantum phase transition as shown in
Fig.~\ref{Fig:PD}.  As we will discuss in Sec.~\ref{Sect:Corr} such
terms also influence the non-universal amplitudes in the correlation
functions whilst preserving the universal critical exponents.

\section{Local Expectation Values}
\label{Sect:Local}

\begin{figure}
  \includegraphics[clip=true,width=3.2in]{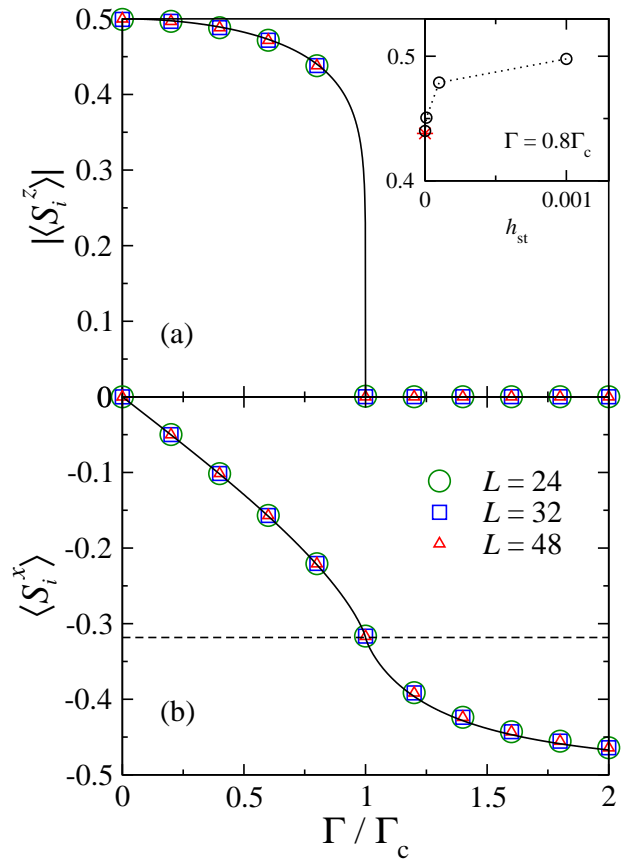}
  \caption{(color online). Local expectation values deep within the Mott phase
    obtained by DMRG for the bosonic Hamiltonian (\ref{HR}) with
    periodic boundaries and restriction parameters $r_m=1$ and
    $r_a=2$. In order to demonstrate the broader validity of our
    results, whilst suppressing higher order terms in the strong
    coupling expansion, we choose different parameters to those used
    in Fig.~\ref{Fig:PD}.  We set $\epsilon_a=0$ and
    $\epsilon_m=U_{aa}=3.8$, corresponding to $h=0$ and take
    $U_{am}=4$ and $t=0.005$. (a) Staggered magnetization $|\langle
    S^z_i\rangle|$.  The theoretical result for the transverse field
    Ising model is given by Eq.~(\ref{TFI:Stagg}) and is indicated by
    the solid line. Inset: The application of a small staggered field
    $h_{st}$ yields the same results as in panel (a) in the limit
    $h_{\rm st}\rightarrow 0$. (b) Transverse magnetization $\langle
    S_i^x\rangle$. The solid line shows the Ising behavior given by
    Eq.~(\ref{TFI:Sx}). The dashed line corresponds to $\langle
    S_i^x\rangle=-1/\pi$ which holds at criticality.}
\label{Fig:Local}
\end{figure}

Having provided numerical evidence for an Ising quantum phase
transition in the second Mott lobe of the Hamiltonian (\ref{atmolham})
we turn our attention to the local expectation values. The order
parameter of the antiferromagnetic transverse field Ising model
(\ref{Ising}) is the staggered magnetization $(-1)^i\langle
S_i^z\rangle$. In the thermodynamic limit with $h=0$ and
$\Gamma<\Gamma_{\rm c}$ this is given by
\cite{Pfeuty:TFI,Hieida:Reentrant}
\begin{equation}
(-1)^i\langle S_i^z\rangle=
\frac{1}{2}\left[1-\left(\frac{\Gamma}{\Gamma_{\rm c}}\right)^2\right]^{\beta},
\label{TFI:Stagg}
\end{equation}
where $\beta=1/8$ is the Ising critical exponent; in the disordered
phase with $\Gamma>\Gamma_{\rm c}$ the order parameter vanishes.  In
order to verify this characteristic behavior in the bosonic
Hamiltonian (\ref{HR}) we choose parameters deep within the Mott
phase.  As shown in Fig.~\ref{Fig:Local}(a)
our DMRG results are in excellent agreement with
Eq.~(\ref{TFI:Stagg}). This confirms that any higher order corrections
to the Ising description (\ref{Ising}) are small. In a similar fashion
the transverse magnetization is given by
\cite{Pfeuty:TFI,Hieida:Reentrant}
\begin{equation}
\langle S_i^x\rangle =-\int_0^\pi \frac{dk}{2\pi}\,
\frac{2\Gamma+J_{zz}\cos k}{\sqrt{4\Gamma^2+J_{zz}^2+4\Gamma J_{zz}\cos k}}.
\label{TFI:Sx}
\end{equation}
This dependence is confirmed in Fig.~\ref{Fig:Local}(b) both at and
away from criticality. Note that at the critical point where
$\Gamma=\Gamma_{\rm c}$ the expectation value $\langle
S_i^x\rangle=-1/\pi$ is non-vanishing due to the transverse field.  We
have also checked the consistency of our DMRG results by applying a
small staggered field $h_{\rm st}$ to the bosonic Hamiltonian
(\ref{HR}), $\Delta H_{{\rm st}}=-h_{\rm st}\sum_i(-1)^iS_i^z$, where
$S_i^z$ is given by Eq.~(\ref{spindef}). In the limit $h_{\rm
  st}\rightarrow 0$ this replicates the effect of spontaneous symmetry
breaking in our finite size simulations. The results converge to the
same values as for $h_{\rm st}=0$; see inset of
Fig.~\ref{Fig:Local}(a).

\begin{figure}
  \includegraphics[clip=true,width=3in]{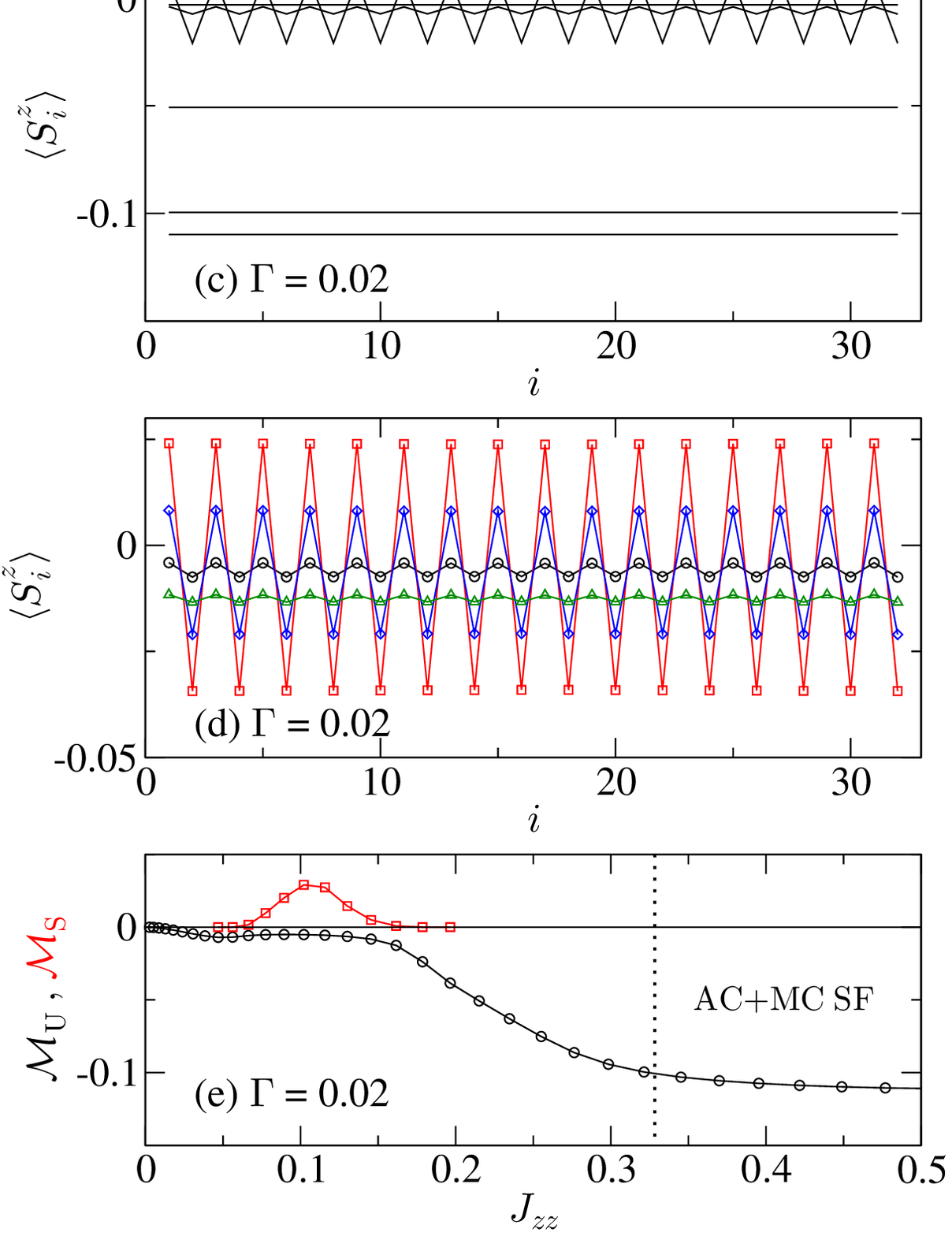}
  \caption{(color online). (a) DMRG results for the local
      magnetization $\langle S^z_i\rangle$ for the parameters used in
      Fig.~\ref{Fig:PD} with $h=0$ and $\Gamma=0.04$. We set $t=0.025,
      0.075, 0.125,..., 0.325$ from top to bottom with $L=32$
      and periodic boundaries.  The plateaus correspond to the
      development of a uniform magnetization whilst the staggered
      magnetization remains zero.  (b) Uniform magnetization
      ${\mathcal M}_{\rm U}=\sum_i \langle S_i^z\rangle/L$. The
      non-zero value is attributed to higher order terms in the strong
      coupling expansion. These may modify the leading order magnetic
      field given in Eq.~(\ref{Isingh}). The dotted line shows the
      approximate location of the MI to AC+MC transition obtained from
      the gap data in Fig.~\ref{Fig:PD}.  (c) We set $\Gamma=0.02$ and
      use the same values of the remaining parameters as in panel
      (a). The oscillations correspond to the onset of the ${\mathbb
        Z}_2 $ ordered phase in Fig.~\ref{Fig:PD}. (d)
      Antiferromagnetic oscillations in the ${\mathbb Z}_2$ ordered
      phase with $t=0.125$ (circles), $t=0.155$ (squares), $t=0.175$
      (diamonds) and $t=0.195$ (triangles). (e) Uniform magnetization
      ${\mathcal M}_{\rm U}$ (circles), and staggered magnetization
      ${\mathcal M}_{\rm S}=\sum_i (-1)^i\langle S_i^z\rangle/L$
      (squares) for $\Gamma=0.02$.}
\label{Fig:Localmag}
\end{figure}

Having established the validity of the explicit Ising Hamiltonian
(\ref{Ising}) deep within the Mott phase, it is instructive to see how
this leading order behavior is modified as one moves out towards the
MI-SF boundary. In Fig.~\ref{Fig:Localmag}(a) we show the evolution of
the local magnetization $\langle S_i^z\rangle$ with increasing hopping
parameters within the ${\mathbb Z}_2$ disordered phase shown in
Fig.~\ref{Fig:PD}.
Deep within the MI, $\langle S^z_i\rangle=0$ as one would expect for
the disordered phase of the transverse field Ising model (\ref{Ising})
with $h=0$. However, at larger values of the hopping parameters a
finite uniform magnetization develops as indicated by the plateaus in
Fig.~\ref{Fig:Localmag}(a). This is due to the presence of higher
order terms in the strong coupling expansion. In general such
contributions are expected to induce corrections to the leading order
coefficients, $J_{zz}$ and $C$, and the magnetic field, $h$, given in
Eqs. (\ref{Isingj}) - (\ref{Isingc}).  However, it is evident from
Fig.~\ref{Fig:Localmag}(b) that this uniform contribution to the
magnetization remains significantly below the saturation value of
$\langle S_i^z\rangle=-1/2$, indicating that the leading order
description (\ref{Ising}) is a useful starting point in the broader
phase diagram. Indeed, the staggered magnetization remains zero in the
bulk of the system, as may be seen from the absence of oscillations in
Fig.~\ref{Fig:Localmag}(a). As such we remain in the disordered phase
of an Ising antiferromagnet, albeit in the presence of an increasing
uniform effective magnetic field.  More generally, one may also
  transit through the ${\mathbb Z}_2$ ordered region shown in
  Fig.~\ref{Fig:PD}. As indicated in Figs.~\ref{Fig:Localmag}(c) and
  (d) this results in the onset of antiferromagnetic oscillations in
  the local magnetization. In Fig.~\ref{Fig:Localmag}(e) we plot the
  evolution of the corresponding uniform and staggered
  magnetizations. The region of support of the staggered component 
is in agreement with the ${\mathbb Z}_2$ ordered phase inferred 
from the gap data in Fig.~\ref{Fig:PD}.

\section{Correlation Functions}
\label{Sect:Corr}
\begin{figure}
  \includegraphics[clip=true,width=3in]{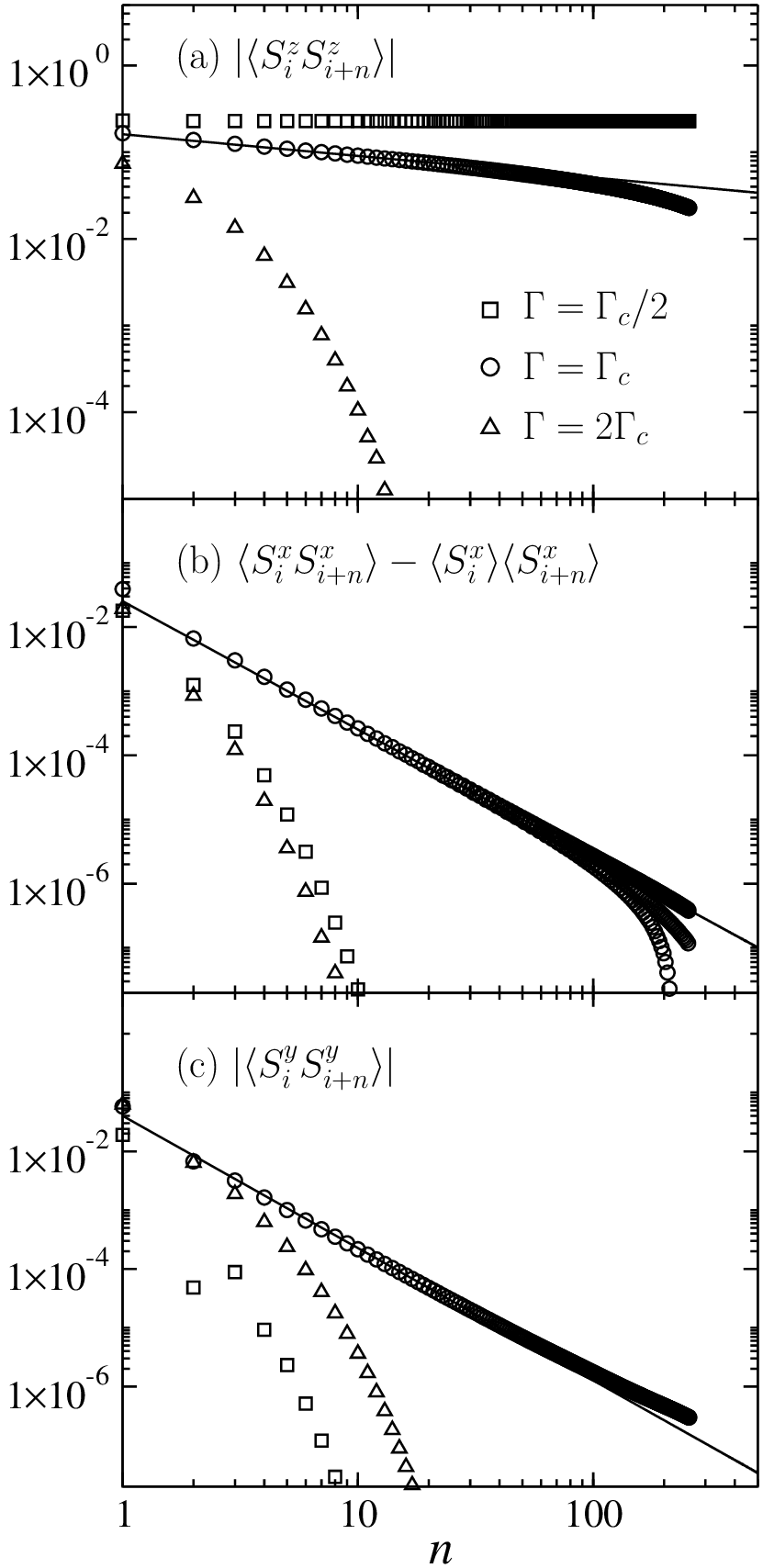}
  \caption{Correlation functions within the second Mott lobe of the
    bosonic Hamiltonian (\ref{HR}) obtained by DMRG with $L=512$ and
    open boundaries.  We use the same parameters as in
    Fig.~\ref{Fig:AFMgap}. The data are consistent with an Ising
    quantum phase transition between a magnetic state with long range
    order and a disordered state with a finite correlation length.
    The solid lines correspond to the critical two-point functions in
    Eqs.~(\ref{zzcrit}), (\ref{xxcrit}) and (\ref{yycrit}). A
    quantitative demonstration of the Ising critical exponents is
    shown in Fig.~\ref{Fig:DMRGcollapse1} using periodic boundaries
    and finite size scaling.}
\label{Fig:Mottcorr}
\end{figure}

\begin{figure}[h!]
  \includegraphics[clip=true,width=3.2in]{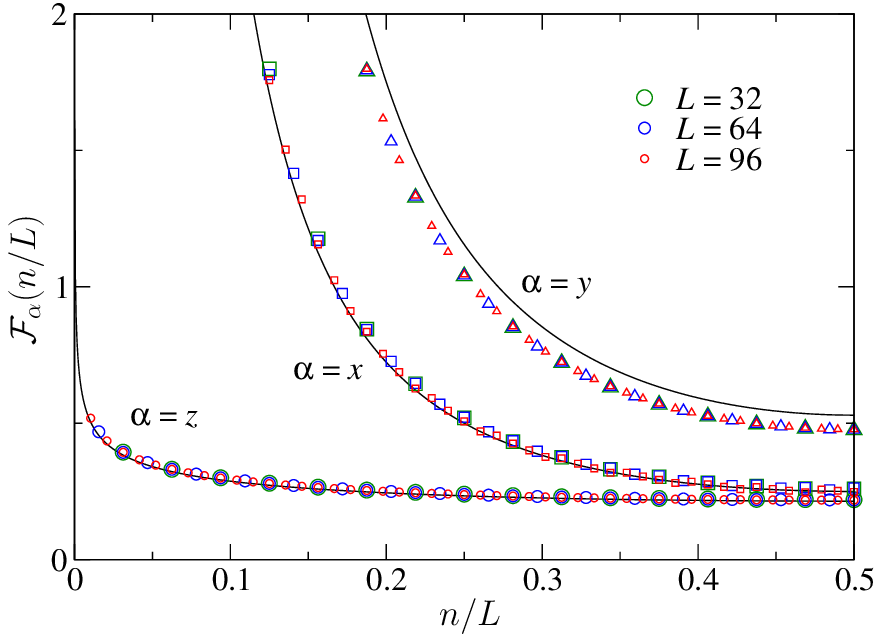}
  \caption{(color online). DMRG results for the rescaled bosonic
    correlation functions $L^{1/4}|\langle S_i^zS_{i+n}^z\rangle_L|$
    (circles), $L^{2}[\langle S_i^xS_{i+n}^x\rangle-\langle
    S_i^x\rangle\langle S_{i+n}^x\rangle]_L$ (squares) and
    $L^{9/4}|\langle S_i^yS_{i+n}^y\rangle_L|$ (triangles) with
    periodic boundaries at criticality for the parameter set of
    Fig.~\ref{Fig:Mottcorr} with $t=0.05$.  The data collapse over the
    entire system length with the Ising critical exponents. The
    non-universal prefactors differ slightly from the theoretical
    predictions of the lattice Ising model (\ref{Ising}) as indicated
    by the solid lines. This is due to the presence of small
    additional XY contributions to the Ising description. By
    descending deeper into the Mott lobe one obtains a complete
    quantitative agreement including the non-universal prefactors as
    shown in Fig.~\ref{Fig:DMRGcollapse2}.}
\label{Fig:DMRGcollapse1}
\end{figure}

\begin{figure}[h!]
  \includegraphics[clip=true,width=3.2in]{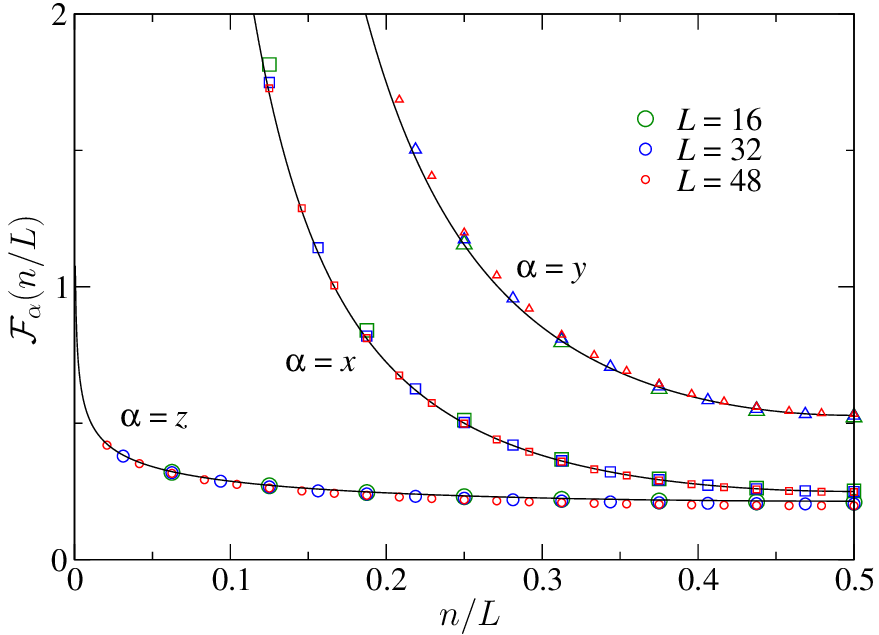}
  \caption{(color online). DMRG results for the rescaled bosonic
    correlation functions $L^{1/4}|\langle S_i^zS_{i+n}^z\rangle_L|$
    (circles), $L^{2}[\langle S_i^xS_{i+n}^x\rangle-\langle
    S_i^x\rangle\langle S_{i+n}^x\rangle]_L$ (squares) and
    $L^{9/4}|\langle S_i^yS_{i+n}^y\rangle_L|$ (triangles) with
    periodic boundaries at criticality for the parameter set of
    Fig.~\ref{Fig:Local} with $t=0.005$. The data show clear scaling
    collapse over the entire system length. The theoretical results
    for the lattice Ising model (\ref{Ising}) in finite size geometry
    are indicated by solid lines. The data are in excellent agreement
    with both the universal Ising critical exponents and the
    non-universal amplitudes in Eqs.~(\ref{zzcrit}), (\ref{xxcrit})
    and (\ref{yycrit}).}
\label{Fig:DMRGcollapse2}
\end{figure}

Having discussed the local expectation values we now consider
correlation functions.  In order to orient the discussion we first
recall the expected theoretical behavior at the critical point of the
antiferromagnetic Ising model in a transverse field, where
$\Gamma_{\rm c}=J_{zz}/2$ and $h=0$. In the absence of a longitudinal
field the asymptotic longitudinal correlation function decays as a
power law \cite{Pfeuty:TFI}
\begin{equation}
  \langle S_i^zS_{i+n}^z\rangle\sim  (-1)^n\,{\mathcal B}\,n^{-\eta}+\dots,
\label{zzcrit}
\end{equation}
where $\eta=1/4$ is the Ising pair correlation exponent, ${\mathcal
  B}={\mathcal A}^{-3}\,4^{-1}\,2^{1/12}\,e^{1/4}\simeq 0.161$ is a
constant prefactor, and ${\mathcal A}\simeq 1.2824$ is the
Glaisher--Kinkelin constant. In a similar fashion, the transverse
correlators also exhibit power law behavior at criticality
\cite{Pfeuty:TFI}
\begin{equation}
\langle S_i^xS_{i+n}^x\rangle \sim \langle S_i^x\rangle^2+(2\pi n)^{-2}+\dots,
\label{xxcrit}
\end{equation}
where $\langle S_i^x\rangle=-1/\pi$ 
at the transition and 
\begin{equation} 
\langle S_i^yS_{i+n}^y\rangle \sim -(-1)^n\,({\mathcal B}/4)\,n^{-9/4}+\dots.
\label{yycrit}
\end{equation}
As suggested by equation (\ref{xxcrit}), due to the finite value of
$\langle S_i^x\rangle=-1/\pi$ at criticality one must consider the 
connected correlation function $\langle
S_i^xS_{i+n}^x\rangle-\langle S_i^x\rangle\langle S_{i+n}^x\rangle$ in
order to see power law behavior.  To establish this dependence in the
bosonic Hamiltonian (\ref{HR}) we perform DMRG calculations
with open boundaries and up to $L=512$ sites. As shown in
Fig.~\ref{Fig:Mottcorr}, 
at $\Gamma=\Gamma_{\rm c}$ the data are consistent with power law
behavior in the bulk of the system away from the sample boundaries. In
addition the data show long range order in $|\langle
S_i^zS_{i+n}^z\rangle|$ for $\Gamma<\Gamma_{\rm c}$, and a finite
correlation length for $\Gamma>\Gamma_{\rm c}$.  Although this
provides direct evidence for a quantum phase transition, a
quantitative determination of the Ising critical exponents requires a
more detailed finite size scaling analysis of the data. This is most
readily achieved using periodic boundary conditions. We recall that in
a finite size critical system with periodic boundary conditions, the
two-point function of a primary field ${\mathcal O}$ may be obtained
by conformal transformation of the strip geometry \cite{Cardy:scal}:
\begin{equation}
\langle {\mathcal O}(r_1){\mathcal O}(r_2)\rangle_L = 
{\mathcal N}\left[\frac{\pi}{L\sin\left(\frac{\pi r}{L}\right)}\right]^{a}.
\label{confcorr}
\end{equation}
Here $a$ is the critical exponent in the thermodynamic limit and
${\mathcal N}$ is the prefactor: $\langle {\mathcal O}(r_1){\mathcal
  O}(r_2)\rangle_\infty={\mathcal N}r^{-a}$. It follows from equation
(\ref{confcorr}) that the rescaled combination 
\begin{equation}
L^a\langle {\mathcal
  O}(r_1){\mathcal O}(r_2)\rangle_L=f_a\left(r/L\right)
\end{equation}
is a prescribed scaling function 
\begin{equation}
  f_a(x)={\mathcal N}\left[\frac{\pi}{\sin(\pi x)}\right]^a,
\end{equation}
of the reduced separation $x=r/L$. As shown in
Figs.~\ref{Fig:DMRGcollapse1} and \ref{Fig:DMRGcollapse2}, the
rescaled critical two-point functions
\begin{equation}
\begin{aligned}
{\mathcal F}_{z}\left(n/L\right) & \equiv L^{1/4}|\langle S_i^zS_{i+n}^z\rangle_L|,\\
{\mathcal F}_{y}\left(n/L\right) & \equiv  L^{9/4}|\langle S_i^yS_{i+n}^y\rangle_L|,\\
{\mathcal F}_{x}\left(n/L\right) & \equiv L^{2}[\langle S_i^x S_{i+n}^x\rangle-\langle S_i^x\rangle \langle
S_{i+n}^x\rangle]_L,
\end{aligned} 
\end{equation}
all show striking data collapse over the entire system length, showing
clear indications of Ising criticality. Deep in the Mott phase the
resulting scaling functions are in excellent agreement with the
theoretical results for the lattice Ising model (\ref{Ising}) in
finite size geometry, as indicated in Fig.~\ref{Fig:DMRGcollapse2}.
This includes both the universal critical exponents, $a$ and the
non-universal amplitude prefactors, ${\mathcal N}$ taken from
Eqs.~(\ref{zzcrit}), (\ref{xxcrit}) and (\ref{yycrit}). Further out in
the Mott phase, the non-universal prefactors are influenced by higher
order terms in the strong coupling expansion as discussed in
Sec.~\ref{Sect:XY}, but the universal Ising exponents are robust to
these perturbations; see Fig.~\ref{Fig:DMRGcollapse1}.

\begin{figure}
  \includegraphics[clip=true,width=3.2in]{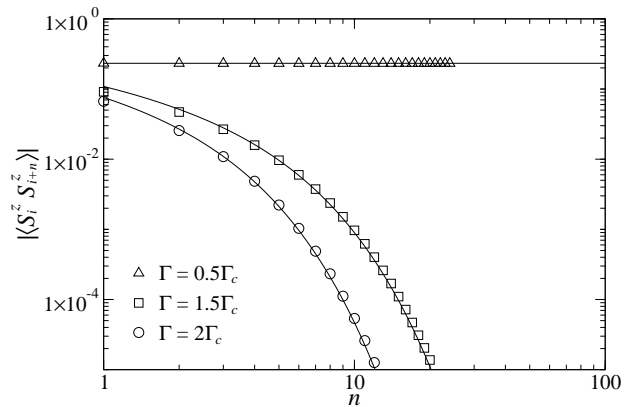}
  \caption{Off-critical order parameter correlations $|\langle
    S_i^zS^z_{i+n}\rangle|$ obtained by DMRG on the 1D system
    (\ref{HR}) with $L=48$ and periodic boundaries.  We use the
    same parameters as in Fig.~\ref{Fig:Local}. The solid lines
    correspond to the quantum Ising model (\ref{Ising}) and are given
    by Eq.~(\ref{zzord}) for $\Gamma<\Gamma_{\rm c}$ and Eq.~(\ref{zzdis})
    for $\Gamma>\Gamma_{\rm c}$. The agreement confirms the presence of long
    range order for $\Gamma<\Gamma_{\rm c}$ and exponential decay for
    $\Gamma>\Gamma_{\rm c}$.}
\label{Fig:Mottzzcorr}
\end{figure}

Having confirmed the presence of a line of critical Ising correlations
within the second Mott lobe of the bosonic Hamiltonian (\ref{HR}) we
now consider the generic behavior in the MI phase. As may be seen in
Fig.~\ref{Fig:Mottcorr}, the data reveal a finite correlation length
for the connected correlation functions on either side of the
transition. This is consistent with the presence of an Ising spectral
gap as shown in Fig.~\ref{Fig:AFMgap}. In order to place this behavior
on a quantitative footing we recall the principal results for the
transverse field Ising model (\ref{Ising}) with $h=0$.  In the ordered
phase with $\Gamma<\Gamma_{\rm c}$ the leading contribution to the
longitudinal correlations is given by \cite{Pfeuty:TFI}
\begin{equation}
  \langle S_i^zS_{i+n}^z\rangle \sim (-1)^n|\langle S_i^z\rangle|^2,
\label{zzord}
\end{equation}
where $|\langle S_i^z\rangle|$ is the staggered magnetization
corresponding to the onset of long range antiferromagnetic order as
given by Eq.~(\ref{TFI:Stagg}).  In contrast, in the disordered phase
with $\Gamma>\Gamma_{\rm c}$, the correlations decay exponentially with a
power law prefactor \cite{Pfeuty:TFI}
\begin{equation}
  \langle S_i^zS_{i+n}^z\rangle \sim \frac{(-1)^n}{4}
  \left[1-\left(\frac{\Gamma_{\rm c}}{\Gamma}\right)^2\right]^{-1/4}
\frac{e^{-n/\xi}}{\sqrt{\pi n}},
\label{zzdis}
\end{equation}
where $\xi^{-1}=\ln(\Gamma/\Gamma_{\rm c})$.  To confirm this behavior
we descend deep in to the Mott phase in order to suppress the effects
of the small XY terms given by Eq.~(\ref{XY}). As shown in
Fig.~\ref{Fig:Mottzzcorr} the results are in excellent agreement with
the theoretical predictions (\ref{zzord}) and (\ref{zzdis}).

\section{Atom--Molecule Correlations}
\label{Sect:Atmol}

The above considerations are consistent with the notion that away from
the Ising transition line the connected correlations decay
exponentially in the MI. As advocated in Ref.~\cite{Bhaseen:Feshising}
the absence of low lying gapless excitations precludes the possibility
of the novel super-Mott behavior proposed in
Ref.~\cite{Rousseau:Fesh,Rousseau:Mixtures}. This gained further
support in a recent comment \cite{Eckholt:Comment} which confirms the
presence of exponential decay in the atom--molecule correlation
functions.  We discuss these observations for the bosonic Hamiltonian
(\ref{HR}) within the framework of the Ising description.

\begin{figure}[!h]
  \includegraphics[clip=true,width=3.2in]{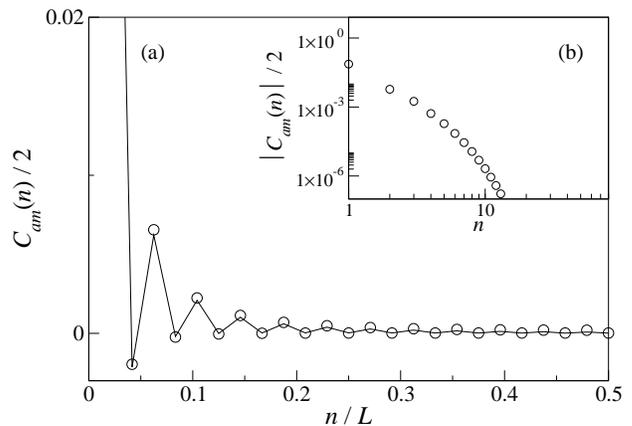}
  \caption{Correlation function $C_{am}(n)/2$ at and away from
    criticality obtained by DMRG on the 1D system (\ref{HR}) with
    $L=48$ and periodic boundaries. We use the same parameters as in
    Fig.~\ref{Fig:Local}. (a) At criticality clear oscillations are
    present in conformity with the Ising description.  The solid line
    corresponds to the theoretical prediction in
    Eq.~(\ref{transcritfinite}).  (b) Away from the critical point
    with $\Gamma=2\Gamma_{\rm c}$, $C_{am}(n)$ shows a finite
    correlation length due to the Ising gap. }
\label{Fig:Transcorr}
\end{figure}

In order to investigate the possibility of counterflow supercurrents
\cite{Kuklov:Counter,Hu:Counter} of atoms and molecules, the authors of
Ref.~\cite{Eckholt:Comment} consider the connected correlation
function
\begin{equation}
\begin{aligned}
  C_{am}(n) & =\langle m^\dagger(n)a(n)a(n)a^\dagger(0)a^\dagger(0)m(0)\rangle\\
  & \hspace{0.5cm} -\langle m^\dagger(n)a(n)a(n)\rangle\langle
  a^\dagger(0)a^\dagger(0)m(0)\rangle.
\label{Camdef}
\end{aligned}
\end{equation}
Using the spin mapping (\ref{spindef})
\cite{Bhaseen:Feshising,Hohenadler:QPT} this may be written in the
equivalent form
\begin{equation}
C_{am}(n)=2\left[\langle S^+(n)S^-(0)\rangle
-\langle S^+(n)\rangle\langle S^-(0)\rangle\right].
\label{cspin}
\end{equation}
Deep within the second Mott lobe one may therefore use the Ising
Hamiltonian (\ref{Ising}) to gain a handle on the bosonic correlation
function (\ref{Camdef}). To this end we may decompose the first term
in Eq.~(\ref{cspin}) as
\begin{equation}
\langle S^+_iS^-_{i+n}\rangle=\langle S_i^xS_{i+n}^x\rangle+\langle
S_i^yS_{i+n}^y\rangle,
\end{equation}
where the mixed component terms cancel. It follows that 
\begin{equation}
  C_{am}(n)=2\left[\langle S_i^xS_{i+n}^x\rangle-\langle S_i^x\rangle \langle S_{i+n}^x\rangle +\langle
    S_i^yS_{i+n}^y\rangle\right],
\end{equation}
where we use the fact that $\langle S_i^y\rangle=0$ for the transverse
field Ising model (\ref{Ising}) in zero magnetic field. That is to
say, $C_{am}(n)$ is the sum of the connected $xx$ correlation function
and the $yy$ correlation function. From our previous discussion in
Sec.~\ref{Sect:Corr} these contributions generically decay
exponentially, in agreement with the findings of
Ref.~\cite{Eckholt:Comment}.  However, along the locus of the Ising
quantum phase transition one expects power law contributions.  Using
Eqs.  (\ref{xxcrit}) and (\ref{yycrit}) one obtains
\begin{equation}
  C_{am}(n)\sim 2\left[(2\pi n)^{-2}-(-1)^n({\mathcal B}/4)n^{-9/4}\right].
\label{transcrit}
\end{equation}
In order to confirm this characteristic oscillatory dependence we
perform DMRG on the 1D bosonic Hamiltonian (\ref{HR}) with
$L=96$ and periodic boundaries.  Employing the 
finite size result (\ref{confcorr}) obtained by conformal transformation 
one obtains
\begin{equation}
\frac{C_{am}(n)}{2}\sim 
\frac{1}{(2\pi)^2}\left[\frac{\pi/L}{\sin\left(\frac{\pi n}{L}\right)}\right]^2
-(-1)^n\frac{{\mathcal B}}{4}\left[\frac{\pi/L}{\sin\left(\frac{\pi n}{L}\right)}\right]^{9/4}.
\label{transcritfinite}
\end{equation}
As shown in Fig.~\ref{Fig:Transcorr}
the results are in excellent agreement with the underlying Ising
correlation functions. 

\section{Canonical Softcore Bosons}
\label{Sect:Soft}

\begin{figure}
  \includegraphics[clip=true,width=3.2in]{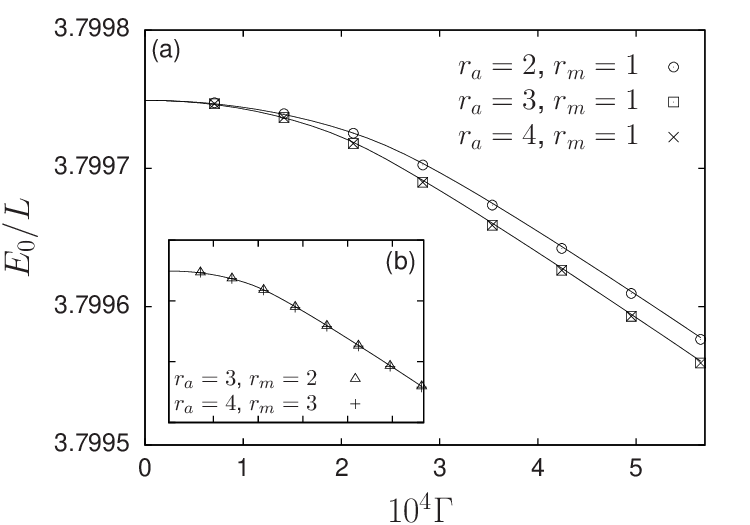}
  \caption{Ground state energy density of the Hamiltonian (\ref{HR})
    obtained by DMRG with $L=20$ (to aid comparison with ED results
    for the Ising model) and periodic boundaries. (a) We use the
    parameters in Fig.~\ref{Fig:Local} and increase the local atomic
    Hilbert space restriction $r_a$, with $r_m=1$ held fixed.  The
    lines are results for the energy density of the Ising Hamiltonian
    (\ref{Ising}) obtained by ED ($L=20$) with $J_{zz}=5.02\times
    10^{-4}$ and $h = 0$ for $r_a=2$, and $J_{zz}=4.23\times 10^{-4}$
    and $h = 7.89\times 10^{-5}$ for $r_a=3,4$. The change from
    $r_a=2$ to $r_a=3$ is due to the presence of additional virtual
    states which modify the Ising model coefficients. The absence of
    any change beyond $r_a=3$ is consistent with second order
    perturbation theory around the second Mott lobe.  (b) Ground
      state energy density for the same parameters as in panel (a)
      with the additional interaction $U_{mm}=4$. We increase the
      local Hilbert space from $r_a=3$ and $r_m=2$ to $r_a=4$ and
      $r_m=3$. The absence of any further change is consistent with
      the maximum occupancy for virtual states explored in second
      order perturbation theory around the second Mott lobe. The solid
      line is the energy density of the Ising Hamiltonian
      (\ref{Ising}) obtained by ED ($L=20$) with $J_{zz}=4.16\times
      10^{-4}$ and $h=7.27\times 10^{-5}$ as given by Eqs.
      (\ref{Jzzsoft}) and (\ref{hsoft}) respectively.}
\label{Fig:evg}
\end{figure}

\begin{figure}
\includegraphics[width=3.2in,clip]{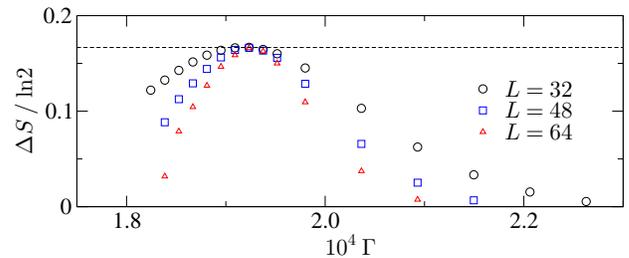}
\caption{Entanglement entropy difference $\Delta S(L)\equiv
    S_L(L/2)-S_{L/2}(L/4)$ showing an Ising quantum phase transition
    within the second Mott lobe for restricted softcore bosons. We
    truncate the local Hilbert space to $r_a=3$ and $r_m=2$ which is
    the maximum occupancy explored at second order in perturbation
    theory. We use the same parameters as in Fig.~\ref{Fig:evg}(b) and
    set $t=0.005$.  The peak height corresponds to a central charge
    $c\approx 1/2$.}
\label{Fig:Deltas}
\end{figure}

Having provided a discussion of the model (\ref{HR}) in the reduced
Hilbert space, with $r_a=2$ and $r_m=1$, we turn our attention to the
more general problem with canonical softcore bosons. In this situation
one must allow for the presence of additional virtual intermediate
states in the magnetic description.  For example, in a configuration
with two pairs of $a$-atoms on neighboring sites the softcore problem
allows virtual hopping to take place, in contrast to the problem with
$r_a=2$.  This modifies the coefficients, $J_{zz}$, $h$, and $C$, but
the Ising description remains valid deep within the second Mott
lobe. One again obtains the effective magnetic Hamiltonian given in Eq.~(\ref{Ising}) but with the modified coefficients,
\begin{equation}
J_{zz}=\frac{4t_a^2}{U_{am}-U_{aa}}+\frac{t_m^2}{U_{am}}
-\frac{12t_a^2}{U_{aa}}
-\frac{4t_m^2}{U_{mm}},
\label{Jzzsoft}
\end{equation}
and 
\begin{equation}
  h=\epsilon_m-2\epsilon_a-U_{aa}+\frac{z}{2}\left(\frac{12t_a^2}{U_{aa}}
    -\frac{4t_m^2}{U_{mm}}\right),
\label{hsoft}
\end{equation}
together with 
\begin{equation}
\begin{aligned}
  C & =  L\left(\frac{\epsilon_m}{2}+\epsilon_a+\frac{U_{aa}}{2}\right. \\
  & \left. -\frac{z}{8}
    \left\{\frac{4t_a^2}{U_{am}-U_{aa}}+\frac{t_m^2}{U_{am}}+
      \frac{12t_a^2}{U_{aa}}
      +\frac{4t_m^2}{U_{mm}}
    \right\}\right),
\label{csoft}
\end{aligned}
\end{equation}
as shown in Appendix \ref{Sect:Effec}. Note in particular that the
effective longitudinal magnetic field, $h$, now depends on the hopping
parameters and is therefore already present at second order in the
strong coupling expansion.  In order to see the effect of these
additional intermediate states it is instructive to examine the change
in the ground state energy of the bosonic Hamiltonian (\ref{HR}) upon
increasing the local atomic Hilbert space restriction, $r_a$. For
simplicity, we consider $U_{mm}\rightarrow\infty$, corresponding to
hardcore molecules with $r_m=1$. As shown in Fig.~\ref{Fig:evg}(a),
the ground state energy changes on going from $r_a=2$ to $r_a=3$ due
to these additional hopping processes.
However, increasing the atomic restriction beyond $r_a=3$ has no
further effect, since higher occupations are not explored at second
order in perturbation theory within the second Mott lobe.  The
excellent agreement of the bosonic results with $r_a=3$ and $r_a=4$
therefore supports the applicability of the second order Ising
description.  The solid lines shown in Fig.~\ref{Fig:evg}(a)
correspond to the ground state energy density of the Ising model
(\ref{Ising}) obtained by exact diagonalization with the appropriate
coefficients.  Our DMRG results are in excellent agreement with these
predictions. As shown in Fig.~\ref{Fig:evg}(b) this agreement
  also extends to the generic softcore problem. At second order in
  perturbation theory around the second Mott lobe the maximum local
  occupation is $r_a=3$ and $r_m=2$. The absence of any further change
  in the ground state energy on increasing the local Hilbert space
  justifies both the use of this truncation and the Ising
  description. In Fig.~\ref{Fig:Deltas} we show the existence of an
  antiferromagnetic Ising quantum phase transition within the second
  Mott lobe for these restricted softcore bosons. The entanglement
  entropy difference as given by Eq.~(\ref{deltas}) exhibits a peak on
  transiting through the magnetic transition. At criticality $\Delta
  S=\frac{c}{3}\ln 2+\dots$
  and the peak height is in good agreement with an Ising quantum phase
  transition with central charge $c=1/2$.
\section{Ferromagnetic Ising Transition}
\label{Sect:FM}

\begin{figure}[!h]
  \includegraphics[clip=true,width=3.2in]{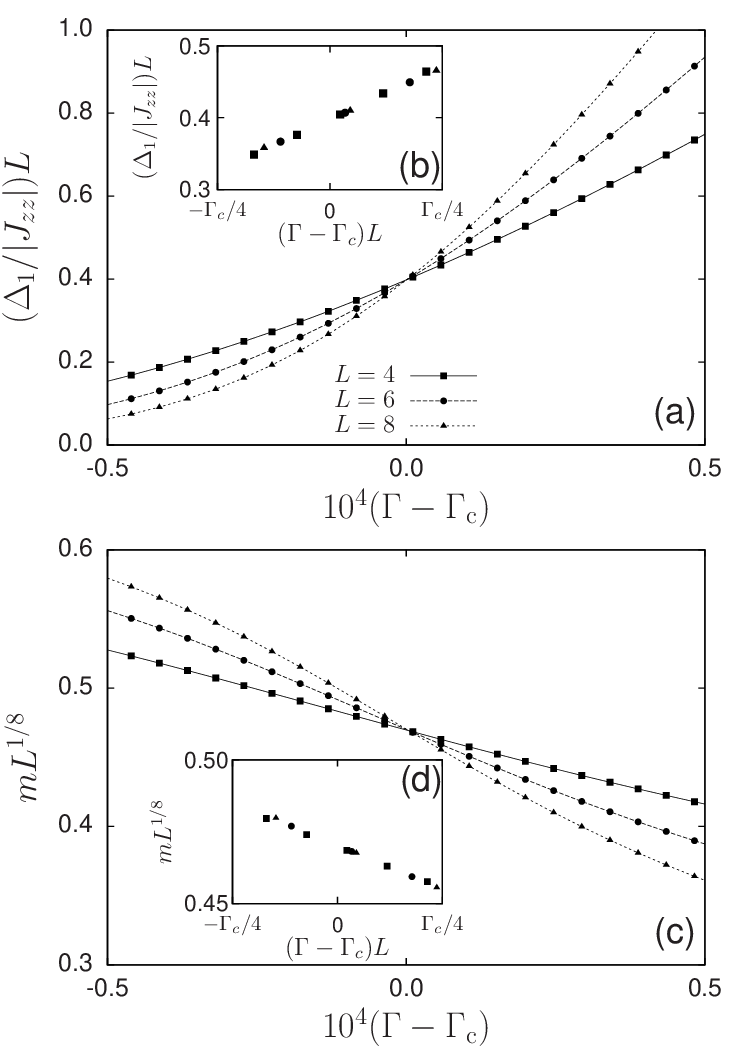}
  \caption{ED results for the Hamiltonian
    (\ref{HR}) with hardcore molecules ($r_m=1$) and up to three
    atoms ($r_a=3$) per site. We set $\epsilon_a=0$, $U_{aa}=2$,
    $U_{am}=4$, $t_a=2t_m$ and take $t_a=0.01$ corresponding to
    FM exchange with $J_{zz}\simeq-3.94\times 10^{-4}$. For
    simplicity we set $h=\epsilon_m-2+6t_a^2=0$ by taking
    $\epsilon_m=1.9994$. (a) The rescaled energy gap $\Delta_1\equiv
    E_1-E_0$ shows a clear intersection at $\Gamma_{\rm c}\approx
    1.969\times 10^{-4}\approx 0.4997 |J_{zz}|$ corresponding to a
    FM transition. (b) Scaling collapse with the
    Ising critical exponent, $\nu=1$. (c) The rescaled
    pseudo-magnetization $m=\langle |\sum_i S_i^z|\rangle/L$ indicates a
    transition at the same value of $\Gamma_{\rm c}$. (d) Scaling collapse
    with the Ising critical exponent $\beta=1/8$.}
\label{Fig:FM}
\end{figure}

\begin{figure}[!h]
  \includegraphics[clip=true,width=3.2in]{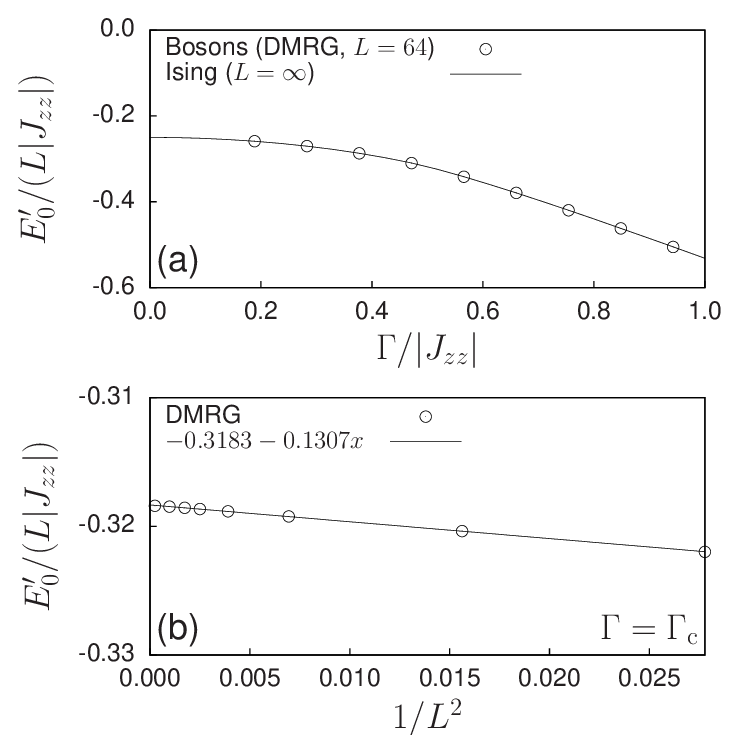}
  \caption{(a) Rescaled ground state energy $E_0^\prime\equiv E_0-C$
    of the bosonic Hamiltonian (\ref{HR}) with the FM parameters
    used in Fig.~\ref{Fig:FM}. The DMRG data are in good agreement
    with Eq.~(\ref{endens}) for the thermodynamic limit of the Ising
    model. (b) The residual finite size corrections,
    $E_0^\prime/L|J_{zz}|=e_\infty/|J_{zz}|-\pi c/12L^2+\dots$, are
    consistent with the central charge $c\approx 0.499$ of an Ising
    transition to a FM state in the absence of a longitudinal magnetic
    field.}
\label{Fig:energycentral}
\end{figure}

An interesting aspect of the canonical softcore result (\ref{Jzzsoft})
is that one may explore both antiferromagnetic and ferromagnetic Ising
interactions due to the relative minus signs. This is readily seen by
exact diagonalization on small systems using the techniques employed
in Refs.~\cite{Bhaseen:Feshising,Hohenadler:QPT,Um:FiniteTFI}.  As
shown in Fig.~\ref{Fig:FM} an Ising transition indeed persists with
ferromagnetic parameters and $h=0$.
This is also confirmed by DMRG results for the ground state energy of
the bosonic Hamiltonian (\ref{HR}) with $L=64$ and periodic boundary
conditions. In the thermodynamic limit, the ground state energy
density of the Ising model (\ref{Ising}) is given by
\cite{Pfeuty:TFI}, $E_0/L=C/L+e_\infty$, where
\begin{equation} e_\infty = -\frac{1}{4\pi}\int_0^\pi dk
  \sqrt{4\Gamma^2+J_{zz}^2+4\Gamma J_{zz}\cos k}.
\label{endens}
\end{equation}
Defining $E_0^\prime\equiv E_0-C$ the result for
$E_0^\prime/L$ is in good agreement with Eq.~(\ref{endens}) as
shown in Fig.~\ref{Fig:energycentral} (a). In addition the finite
size corrections to the ground state energy are well described by the
conformal result \cite{Blote:Conformal,Affleck:Universal} 
\begin{equation}
\frac{E_0^\prime}{L}=e_\infty-\frac{\pi cv}{6L^2}+\dots,
\end{equation}
 where $v=|J_{zz}|/2$ is the
effective velocity and $c$ is the central charge.
Going beyond our $h=0$ parameter choice, this opens up the exciting
possibility of studying the celebrated ${\rm E}_8$ mass spectrum of
the ferromagnetic Ising model in a magnetic field
\cite{Zamolodchikov:Integrals}. As exemplified by recent experiments
on the quasi-one-dimensional Ising ferromagnet ${\rm Co}{\rm Nb}_2{\rm
  O}_6$ (cobalt niobate) \cite{Coldea:E8,Moore:E8}, it would be
interesting to probe the dynamical correlation functions of the 1D
bosonic Hamiltonian (\ref{atmolham}) in more detail. The non-trivial
excitations will manifest themselves at specific frequencies dictated
by the emergent ${\rm E}_8$ mass spectrum. Similar behavior is also
expected in the AFM case in the presence of a staggered longitudinal
magnetic field.

\section{Conclusions} 
\label{Sect:Conc}

We have investigated the Mott insulating state of bosonic pairing
Hamiltonians using analytical and numerical techniques. We have
described the behavior of a broad range of physical observables,
including local expectation values and correlation functions, within
the framework of the paradigmatic quantum Ising model. As advocated in
Refs.~\cite{Bhaseen:Feshising,Hohenadler:QPT,Eckholt:Comment} our
results are consistent with the absence of super--Mott behavior within
the second Mott lobe. Our results differ from the usual two-band
  Bose--Hubbard model which exhibits counterflow superfluidity
  \cite{Kuklov:Counter,Hu:Counter} since Feshbach resonant pairing
  interactions favor a distinct Mott phase with either two atoms or
  one molecule per site. As such, XY terms generically appear at cubic
  order in the hopping strengths and are intrinsically suppressed.  An
  alternative way to see this is to note that for finite Feshbach
  coupling, the symmetry of the Hamiltonian is reduced from ${\rm
    U}(1)\times {\rm U}(1)$ down to ${\rm U}(1)\times{\mathbb
    Z}_2$. As such, one naively expects Ising transitions to occur in
  a Mott phase with large phase fluctuations and an unbroken ${\rm
    U}(1)$ symmetry. Nonetheless, one cannot rule out the possibility
  of novel transitions in other regions of the phase diagram due to
  higher order terms in the strong coupling expansion becoming
  appreciable. It would be interesting to explore this in more
  detail. In addition, we highlight the possibility of using these
systems to explore the ${\rm E}_8$ mass spectrum of the FM Ising model
in a magnetic field. There are many directions for further research
including the influence of higher bands and the possibility of
realizing Ising transitions in Bose--Fermi mixtures; see Appendix
\ref{Sect:Fermi}.

\begin{acknowledgments}
  We are grateful to F. Assaad, J. Oitmaa, F. Pollmann, N. Shannon and
  A. Tsvelik for helpful discussions. MJB, AOS and BDS acknowledge
  EPSRC grant no.  EP/E018130/1. FHLE by EP/D050952/1.  SE and HF
  acknowledge funding by the DFG through grant SFB 652. MH by DFG
  FG1162.
\end{acknowledgments} 

\appendix

\section{Effective Magnetic Hamiltonian}
\label{Sect:Effec}
To obtain the effective spin description 
within the second Mott lobe we perform degenerate perturbation theory. We 
partition the Hamiltonian as $H=H_0+V$ where
\begin{equation}
H_{0} =\sum_{i} \left[H_{i}^{\prime}(1-\mathcal{P}_{0i})+(\epsilon_{m}-h_{0}/2)\mathcal{P}_{0i}\right],
\end{equation}
and
\begin{equation}
V =-\sum_{\expec{ij}\alpha}t_{\alpha}(b^{\dagger}_{i\alpha}b_{j\alpha}+\text{H.c.})+H_{{\rm F}}+h_{0}\sum_{i}S^{z}_{i}\mathcal{P}_{0i}.
\end{equation}
Here $h_{0}=\epsilon_{m}-2\epsilon_{a}-U_{aa}$, $\mathcal{P}_{0i}=\ket{\Uparrow}_{i}\bra{\Uparrow}_{i}+\ket{\Downarrow}_{i}\bra{\Downarrow}_{i}$
and
\begin{equation}
H^{\prime}_{i} =\sum_{\alpha}\epsilon_{\alpha}n_{i\alpha}+\sum_{\alpha\alpha^{\prime}}\frac{U_{\alpha\alpha^{\prime}}}{2}:\!n_{i\alpha} n_{i\alpha^{\prime}}\!:.
\end{equation}
This approach is appropriate deep in the Mott lobe and in the vicinity
of the transition where $g$, $h_{0}$ and $t^{2}/U$ are small and $V$
may be treated perturbatively. Up to second order in $V$ the effective
Hamiltonian is
\begin{equation}
H_{\text{eff}}=\mathcal{P}_{0}\left(H_{0}+V+V(1-\mathcal{P}_{0})\frac{1}{E_{0}-H_{0}}(1-\mathcal{P}_{0})V\right)\mathcal{P}_{0}
\end{equation}
where $E_{0}=L(\epsilon_{m}-h_{0}/2)$ and
$\mathcal{P}_{0}=\prod_{i}\mathcal{P}_{0i}$. Since the hopping terms
are the only source of coupling between the degenerate subspace
spanned by $\mathcal{P}_{0}$ and the remaining Hilbert space, the
second order term reduces to
\begin{equation}
  H^{(2)}_{\text{eff}}=\sum_{\expec{ij}\alpha}t^{2}_{\alpha}\mathcal{P}_{0}b^{\dagger}_{i\alpha}b_{j\alpha}\frac{1}{E_{0}-H_{0}}b^{\dagger}_{j\alpha}b_{i\alpha}\mathcal{P}_{0}+\text{{H.c.}}.
\end{equation}
This acts on two sites so within the degenerate subspace
$\mathcal{P}_{0}$, we must consider its action on four basis states.
The action of $H^{(2)}_{\text{eff}}$ on sites with neighboring
molecules yields
\begin{equation}
H^{(2)}_{\text{eff}}\ket{\Uparrow}_{i}\ket{\Uparrow}_{j}=-\frac{4t^{2}_{m}}{U_{mm}+h_{0}}\ket{\Uparrow}_{i}\ket{\Uparrow}_{j},
\label{eq:doublemolocc}
\end{equation}
and for an arrangement of neighboring atoms gives
\begin{equation}
H^{(2)}_{\text{eff}}\ket{\Downarrow}_{i}\ket{\Downarrow}_{j}=-\frac{12t^{2}_{a}}{U_{aa}-h_{0}}\ket{\Downarrow}_{i}\ket{\Downarrow}_{j}.
\label{eq:tripleocc}
\end{equation}
When the neighboring species are different
\begin{equation}
H^{(2)}_{\text{eff}}\ket{\Uparrow}_{i}\ket{\Downarrow}_{j}=-\left(\frac{2t_{a}^{2}}{U_{am}-U_{aa}}+\frac{t_{m}^{2}}{2U_{am}}\right)\ket{\Uparrow}_{i}\ket{\Downarrow}_{j}.
\end{equation}
In order to obtain the second order contribution to the effective
Hamiltonian we collate these terms. In addition we exploit the spin
projection identities
$\ket{\Uparrow}_{i}\bra{\Uparrow}_{i}\mathcal{P}_{0}=(1/2+S^{z}_{i})\mathcal{P}_{0}$
and
$\ket{\Downarrow}_{i}\bra{\Downarrow}_{i}\mathcal{P}_{0}=(1/2-S^{z}_{i})\mathcal{P}_{0}$
and expand the resulting expression to leading (zeroth) order in
$h_{0}$:
\begin{equation}
  H^{(2)}_{\text{eff}}\simeq\mathcal{P}_{0}\left(J_{zz}\sum_{\expec{ij}}S^{z}_{i}S^{z}_{j}+h_{2}\sum_{i}S^{z}_{i}+C_{2}\right)\mathcal{P}_{0},
\end{equation}
where $J_{zz}$ is given by Eq.~(\ref{Jzzsoft})  and 
\begin{equation}
h_{2}=\frac{z}{2}\left(\frac{12t^{2}_{a}}{U_{aa}}-\frac{4t^{2}_{m}}{U_{mm}}\right).
\end{equation}
The constant offset is given by 
\begin{equation}
C_{2}=-\frac{zL}{8}\left(\frac{4t_{a}^{2}}{U_{am}-U_{aa}}+\frac{t_{m}^{2}}{U_{am}}+\frac{12t^{2}_{a}}{U_{aa}}+\frac{4t^{2}_{m}}{U_{mm}}\right).
\end{equation}
Inclusion of the zeroth and first order contributions in $V$ yields the
effective Ising Hamiltonian (\ref{Ising}) with $h=h_0+h_2$ and
$C=E_0+C_2$ as given by Eqs.~(\ref{hsoft}) and
(\ref{csoft}).  In order to obtain the effective Hamiltonian with
$r_{a}=2$ one must exclude the contribution from
Eq.~(\ref{eq:tripleocc}) as the intermediate states involved in the
perturbation do not satisfy the imposed constraint. Similarly if
$r_{m}=1$ the contribution from Eq.~(\ref{eq:doublemolocc}) must be
excluded.

\section{Bose--Fermi Hamiltonian}

\label{Sect:Fermi}

Throughout this manuscript we have focused exclusively on the bosonic
homonuclear Hamiltonian (\ref{atmolham}) and the associated Ising
description. However, it is evident from the general setup shown in
Fig.~\ref{Fig:Secondlobe} that similar results may also emerge with
more than one atomic species. For example, this is confirmed in
Refs.~\cite{Bhaseen:Feshising,Hohenadler:QPT} for the heteronuclear
bosonic case. In this appendix we note that a similar Ising
description may also apply with two-component fermionic atoms and
bosonic molecules. We consider the Bose--Fermi Hamiltonian
\begin{equation}
\begin{split}
H & =\sum_{i\alpha} \epsilon_{\alpha}n_{i\alpha}-\sum_{\langle ij\rangle}\sum_\alpha
t_\alpha(d_{i\alpha}^\dagger d_{j\alpha}+{\rm H.c.})+H_{\rm F}\\
& +\sum_{i,\alpha\neq \alpha^\prime} \frac{U_{\alpha\alpha^\prime}}{2}n_{i\alpha}n_{i\alpha^\prime}+\sum_i\frac{U_{mm}}{2}n_{im}(n_{im}-1),
\label{HBF}
\end{split}
\end{equation}
where $\alpha=\downarrow,\uparrow$ are fermionic atoms, $\alpha=m$ is
a bosonic molecule, $n_{i\alpha}=d_{i\alpha}^\dagger d_{i\alpha}$ and
$H_{\rm F}=g\sum_i(d_{im}^\dagger d_{i\downarrow}d_{i\uparrow}+{\rm
  H.c.})$. Assuming the existence of a second Mott lobe as depicted in
Fig.~\ref{Fig:Secondlobe}, with either two fermionic atoms or a
bosonic molecule per site, one again obtains an Ising Hamiltonian
(\ref{Ising}) acting on the states $|\!\Downarrow\rangle =
d_{\uparrow}^\dagger d_{\downarrow}^\dagger |0\rangle$ and $|\!\Uparrow
\rangle = d_m^\dagger|0\rangle$. The parameters in Eq.~(\ref{Ising})
are now given by $\Gamma=2g$ and
\begin{equation}
\frac{J_{zz}}{2}=\frac{t^{2}_{m}}{U_{\uparrow m}+U_{\downarrow m}}
+\frac{t^{2}_{\downarrow}}{U_{\downarrow m}-U_{\uparrow\downarrow}}
+\frac{t^{2}_{\uparrow}}{U_{\uparrow m}-U_{\uparrow\downarrow}}-\frac{2t^{2}_{m}}{U_{mm}}.
\label{Jf}
\end{equation}
The effective magnetic field is given by
\begin{equation}
  h=\epsilon_m-
  (\epsilon_\downarrow+\epsilon_\uparrow+U_{\uparrow\downarrow})-\frac{2zt_m^2}{U_{mm}}, 
  \label{hf}
\end{equation}
and the constant offset is
\begin{equation}
C=L\left(\epsilon_{m}-\frac{h}{2}+\frac{zJ}{8}-\frac{zt^{2}_{m}}{U_{mm}}\right).
\label{Cf}
\end{equation}
In the limit $U_{mm}\rightarrow\infty$ the results (\ref{Jf}),
(\ref{hf}) and (\ref{Cf}) coincide with the results of Sec.~\ref{Sect:MI2} where we identify $U_{\uparrow\downarrow}=U_{aa}$ and
$U_{\downarrow m}=U_{\uparrow m}=U_{am}$. In view of our findings in
the bosonic problems it would be interesting to examine 
the Bose--Fermi mixture (\ref{HBF}) in more detail.

\end{document}